\documentclass{article}
\begin{document}
\newcommand{\be}{\begin{equation}}
\newcommand{\ee}{\end{equation}}
\newcommand{\bea}{\begin{eqnarray}}
\newcommand{\eea}{\end{eqnarray}}
\newcommand{\nn}{\nonumber \\}
\newcommand{\non}{\nonumber}
\newcommand{\fig}{Fig.\ \ref}
\newcommand{\tab}{Table\ \ref}
\newcommand{\sect}{Section\ \ref}
\newcommand{\eqn}{Eq.\ \ref}
\newcommand{\sgn}{\ensuremath{\mathrm{sgn}}}
\newcommand{\xiabc}{{\scriptscriptstyle \xi=\{\alpha,\beta,\gamma\}}}
                                                                                
\newcommand{\rmin}{\ensuremath{r_{\mathrm{min}}}}
\newcommand{\hmax}{\ensuremath{h_{\mathrm{max}}}}

\centerline{\bf \huge The Romance of the Ising Model}

\vspace{,2in}

\centerline{\bf \Large Barry M. McCoy}
\centerline{\bf \Large State University of New York}
\centerline{\bf \Large Stony Brook, NY}
\vspace{.2in}

\centerline{\bf Abstract}

The essence of romance is mystery. In this talk, given in
honor of the 60th birthday of Michio Jimbo, I
will explore the meaning of this for the Ising model beginning in 1946 with 
Bruria Kaufman and Willis Lamb, continuing with the wedding by 
Jimbo and Miwa in 1980 of the Ising model with the Painlev{\'e} VI 
equation which had been first discovered by Picard in 1889. 
I will conclude with the current fascination of the magnetic susceptibility and
explore some of the mysteries still outstanding.

\section{Introduction}

A search of Google books reveals that the observation

\vspace{.1in}

The essence of romance is mystery 

\vspace{.1in}

\noindent has been made 
by many authors  in many different ways and in many
different contexts ranging from the literary to the scientific. 
But in all contexts romance betokens fascination and the Ising model 
has fascinated many people, including myself, for many decades and 
in spite of many breakthroughs and moments of understanding the mystery
continues to this day. In this talk I will present some of the
milestones of this romance.

\section{Kaufman and Lamb}

In his talk ``The Ising model in two dimensions'' \cite{onsager1} 
presented at the fifth
Battelle Colloquium on Materials Science, held in Geneva and Gstaad,
Switzerland, September 7-12, 1970, Lars Onsager wrote, following a
discussion of his famous 1944 computation  of the free energy \cite{onsager2} 
and a sketch of his 1945
proof of his conjectured spectrum of the transfer matrix,

\vspace{.2in}
``Before long, however, Bruria Kaufman had developed a much better
strategy.

At Columbia University she first asked Willis E. Lamb to direct her
work on order-disorder problems; but he was much too heavily engaged
in an experimental effort, and I was asked to assume the
responsibility. Unable to talk her out of the idea I suggested that
she explore $\cdots$ By the summer of 1946 she had a beautifully
compact computation of the partition function, bypassing all tedious
detail.

By itself that was only a more elegant derivation of an old result but
the approach looked powerful enough to produce a few more new
ones. Very well, how about correlations?'' 

\vspace{.2in}

The history of the Ising model from that time forth has been the study
of these correlations.

But the deeper meaning of this passage from Onsager's paper completely
escaped me until many years later Rodney Baxter wrote to me concerning
a typescript \cite{baxter} that had been given to him which is 
certainly a draft of
Onsager and Kaufman's calculation of the spontaneous magnetization of
the Ising model. Why in the world would Kaufman, who was
creating pioneering mathematics, ask Lamb, an experimental physicist, to
supervise her research? This question was brought into sharp 
focus when Baxter told me that he was going to contact her about 
the authorship of the typescript. She was then living in Tucson, 
Arizona with her husband, Willis Lamb.

So this is the first romance concerned with the Ising model. Both
Bruria and Willis were married to other people in 1946 when Bruria 
asked Willis to be
her research supervisor and he turned her down. But decades later, 
when Kaufman's husband died in 1992, Lamb invited her to Tucson as a
Visiting Scholar at the University of Arizona where he was a
professor. In 1996, after his wife died, Willis and Bruria were married.   

\section{Correlations and form factors}

The great understanding of Kaufman was that the Ising partition
function could be written by use of fermionic methods as the sum of
four Pfaffians \cite{kauf} and that this fermionic method is powerful enough to
write all correlation functions of the Ising model as determinants \cite{ko}.

The Ising model is a system of ``spins''
$\sigma_{j,k}$ at row $j$ and column $k$ of a square lattice which
take on the values $\sigma_{j,k}=\pm 1$ and interact with their
nearest neighbors with the interaction energy
\begin{equation}
{\mathcal E}=-\sum_{j=-L^v}^{L^v}\sum_{k=-L^h}^{L^h}
\{E^h\sigma_{j,k}\sigma_{j+1,k}+
E^v\sigma_{j,k}\sigma_{j+1,k}\}.
\end{equation}
The correlation functions studied by Kaufman and Onsager are
defined as
\begin{equation}
\langle\sigma_{0,0}\sigma_{M,N}\rangle=\lim_{L^v,L^h\rightarrow
  \infty}Z_{L^v,L^h}^{-1}
\sum_{\sigma=\pm1}
\sigma_{0,0}\sigma_{M,N}e^{-{\mathcal E}/k_BT}
\end{equation}
where $T$ is the temperature, $k_B$ is Boltzmann's constant, 
\begin{equation}
Z_{L^v,L^h}=\sum_{\sigma=\pm 1}e^{-{\mathcal E}/k_BT}
\end{equation}
is the partition function and the sum $\sum_{\sigma=\pm1}$ is over all
values of the variables $\sigma_{j,k}$. 

The discovery of Kaufman and Onsager
\cite{ko}
is that the row and diagonal correlations can be written as a sum of
two determinants. These are further simplified by Montroll,
Potts and Ward \cite{mpw} to a single determinant. 
The diagonal  $\langle\sigma_{0,0}\sigma_{N,N}\rangle$ 
and the row  correlations $\langle\sigma_{0,0}\sigma_{0,N}\rangle$ 
can both be written as $N\times N$ Toeplitz determinants 
\begin{equation}
D_N=
\begin{array}{|llll|}
{a}_0&{a}_{-1}&\cdots&{a}_{-N+1}\\
{a}_1&{ a}_0&\cdots&{a}_{-N+2}\\
\vdots&\vdots&&\vdots\\
{a}_{N-1}&{a}_{N-2}&\cdots&{a}_0
\label{dn}
\end{array}
\label{detdn}
\end{equation}
where
\begin{equation}
a_n={1\over 2\pi}\int_{0}^{2\pi}d\theta
e^{-in\theta}\phi(\theta)
\label{an}
\end{equation}
with
\begin{equation}
\phi(\theta)=\left[ {(1-\alpha_1e^{i\theta})(1-\alpha_2e^{-i\theta})\over
(1-\alpha_1e^{-i\theta})(1-\alpha_2e^{i\theta})}\right]^{1/2}.
\label{kern}
\end{equation}
For $\langle\sigma_{0,0}\sigma_{N,N}\rangle$
\begin{equation}
\alpha_1=0,~~~\alpha_2=(\sinh 2E^v/k_BT \sinh 2E^h/k_BT)^{-1}
\end{equation}
and for $\langle \sigma_{0,0}\sigma_{0,N}\rangle$
\begin{equation}
\alpha_1=e^{-2E^v/k_BT}\tanh E^h/k_BT,~~~\alpha_2=e^{-2E^v/k_BT}\coth E^h/k_BT
\end{equation}
and the square roots are defined to be positive at $\theta=\pi.$
These determinants are very efficient for the calculation of the
correlations when $N$ is small. 

However, when $N$ is large the determinental representation
(\ref{detdn} is not an efficient method of calculation and a
different representation must be found. 

The first step in finding this new representation is the computation
of the limiting value as $N\rightarrow \infty$
\begin{equation}
\lim_{N\rightarrow \infty}\langle\sigma_{0,0}\sigma_{0,N}\rangle
=\lim_{N\rightarrow \infty}\langle\sigma_{0,0}\sigma_{N,N}\rangle=
 (1-t)^{1/4}
\label{smag}
\end{equation}
with 
\begin{equation}
t=(\sinh 2E^v/k_BT\sinh 2E^h/k_BT)^{-2},
\label{tm}
\end{equation}
which is valid for $0\leq t \leq 1 $. For $t>1$ the limit vanishes. The 
value of $T$ for which $t=1$ is called the critical temperature $T_c$.
It is the evaluation of this limit for
$\langle\sigma_{0,0}\sigma_{N,N}\rangle$  which is accomplished
by Kaufman and Onsager in the manuscript recently published by Baxter
\cite{baxter}.

The next step in the evaluation of the long distance behavior of the
correlations was made in 1966 by Wu \cite{wu} who computed the first
correction $f^{(2)}_{0,N}$ to (\ref{smag}) as $N\rightarrow \infty$  
for $\langle\sigma_{0,0}\sigma_{0,N}\rangle$ for $T<T_c$ as a
two-dimensional integral and the leading behavior $f^{(1)}_{0,N}$ 
as $N\rightarrow \infty$ of $\langle\sigma_{0,0}\sigma_{0,N}\rangle$ 
for $T>T_c$ as a one-dimensional
integral. These are the first terms in what is now called the form
factor expansion of the correlation functions, which for general $M,N$
is written for $T<T_c$ as
\begin{equation}
\langle\sigma_{0,0}\sigma_{M,N}\rangle
=(1-t)^{1/4}\{1+\sum_{n=1}^{\infty}f^{(2n)}_{M,N}\}
\label{ffm}
\end{equation}
and for $T>T_c$ as
\begin{equation}
\langle\sigma_{0,0}\sigma_{M,N}\rangle
=(1-t)^{1/4}\sum_{n=0}^{\infty}f^{(2n+1)}_{M,N},
\label{ffp}
\end{equation}
where for $T>T_c$ we use the definition
\begin{equation}
t=(\sinh 2E^v/k_BT\sinh 2E^h/k_BT)^2.
\label{tp}
\end{equation}

The derivation of the complete expansions (\ref{ffm}) and (\ref{ffp})
has its own interesting story. In 1976 Wu, McCoy, Tracy and Barouch
\cite{mccoy3} derived an expansion valid for all $N$ of the
correlations in the form for $T<T_c$ of
\begin{equation}
\langle\sigma_{0,0}\sigma_{M,N}\rangle=(1-t)^{1/4}\exp
\sum_{n=0}^{\infty}F^{(2n)}_{M,N}
\label{eformm}
\end{equation}
and for $T>T_c$
\begin{equation}
\langle\sigma_{0,0}\sigma_{M,N}\rangle
=(1-t)^{1/4}\sum_{n=0}^{\infty}G^{(2n+1)}_{M,N}\exp
\sum_{n=0}^{\infty}{\tilde F}^{(2n)}_{M,N}
\label{eformp}
\end{equation}
where $F^{(2n)}_{M,N}$ and ${\tilde F}^{(2n)}_{M,N}$ are $4n$ dimensional
integrals and $G^{(2n+1)}_{M,N}$ are $4n+2$ dimensional integrals.
For all three functions half of the integrals may be executed by
closing a contour integral on a pole. The forms (\ref{eformm}) and
(\ref{eformp}) of the correlation functions are called the exponential forms.

The form factor expansions (\ref{ffm}) and  (\ref{ffp}) are 
obtained from the exponential forms (\ref{eformm}) and (\ref{eformp})
by expanding the exponentials. For a few low values of $n$ this was
done in \cite{mccoy3} in connection with the study of the magnetic 
susceptibility but the general results for the $f^{(n)}_{M,N}$ were
not given by Nickel \cite{ni1} and \cite{ni2} until 1999 and 2000.
 
A curious feature of the derivation given in \cite{mccoy3} of
(\ref{eformm}) and (\ref{eformp})  is that the method of \cite{wu} developed
for the row correlation $\langle \sigma_{0,0}\sigma_{0,N}\rangle$ is not
used; instead the method used by Cheng and Wu \cite{cw} in the study of 
the leading terms of large
separation  behavior of the general correlation $\langle
\sigma_{0,0}\sigma_{M,N}\rangle$ is used. The original method \cite{wu}
of Wu as applied to the correlations $\langle\sigma_{0,0}\sigma_{0,N}\rangle$
and $\langle \sigma_{0,0}\sigma_{N,N}\rangle$ 
was  extended to all orders in 2007 by Lyberg and McCoy
\cite{mccoy5}. The results in \cite{mccoy5}  for the diagonal
form factors $f^{(n)}_{N,N}(t)$ are for $T<T_c$
\begin{eqnarray}
&&f^{(2n)}_{N,N}(t) =\frac{t^{n(N+n)}}{(n!)^2\,\pi^{2n}}\,
\int_0^1\, \prod_{k=1}^{2n}\,dx_k\,\, x_k^N\,\prod_{j=1}^n\,
\left(\frac{(1-tx_{2j})(x_{2j}^{-1}-1)}{(1-tx_{2j-1})
(x_{2j-1}^{-1}-1)}\right)^{1/2}
\nonumber\\
&&\prod_{1\leq j \leq n}\,\prod_{1\leq k \leq n}
\,\left(\frac{1}{1-tx_{2k-1}x_{2j}}\right)^2
\,\prod_{1\leq j<k \leq n}(x_{2j-1}-x_{2k-1})^2\,(x_{2j}-x_{2k})^2,
\nonumber\\
\label{fdm}
\end{eqnarray}
and for $T\, > \, T_c$
\begin{eqnarray}
&&f^{(2n+1)}_{N,N}(t) = \nonumber\\
&&\frac{t^{(n+1/2)N+n(n+1)}}{n!(n+1)!\pi^{2n+1}}
\,\int_0^1\,\prod_{k=1}^{2n+1}\,dx_k\,\,x_k^N
\prod_{j=1}^{n+1}\,x_{2j-1}^{-1}[(1-tx_{2j-1})(x^{-1}_{2j-1}-1)]^{-1/2}
\nonumber\\
&&\prod_{j=1}^{n}\,x_{2j}[(1-tx_{2j})\,
(x^{-1}_{2j}-1)]^{1/2}\, \prod_{1\leq j \leq n+1}\, \prod_{1\leq k \leq n}
\left(\frac{1}{1-tx_{2j-1}x_{2k}}\right)^2
\nonumber\\
&&\prod_{1\leq j <k\leq n+1}\, (x_{2j-1}-x_{2k-1})^2 \, 
\prod_{1\leq j <k \leq n}\,(x_{2j}-x_{2k})^2.
\label{fdp}
\end{eqnarray}
A closely related form for  the row form factor $f^{(n)}_{0,N}$ 
is also obtained in \cite{mccoy5}. 
The results (\ref{fdm}) and (\ref{fdp}) 
have the startling feature that in the diagonal case the
$f^{(n)}_{N,N}$ do not manifestly reduce term by term to the
corresponding functions obtained from \cite{mccoy3}. The
reconciliation of these two forms is one of the
present mysteries of the Ising model.

These diagonal form factor integrals, which on the surface may appear
to be indigestible, have proven to have many very special properties. 

1). All the integrals in (\ref{fdm}) and (\ref{fdp}) reduce at $t=0$ to a
product of two special cases of the celebrated Selberg integral \cite{sel}
\begin{equation}
\int_0^1\cdots \int_0^1\prod_{i=1}^nt_i^{\alpha-1}(1-t_i)^{\beta-1}
\prod_{1\leq i<j\leq n}|t_i-t_j|^{2\gamma}dt_1\cdots dt_n.
\end{equation} 

2) In \cite{mccoy4} it was discovered by Maple calculations the
   $f^{(n)}_{N,N}$ satisfy Fuchsian differential equation with a
   factorized ``Russian doll'' structure
\begin{eqnarray}
&&F_{2n}f^{2n}_{N,N}=0~~{\rm with}~~F_{2n}=L_{2n+1}(N)\cdots
L_3(N)\cdot L_1(N)\\
&&F_{2n+1}f^{(2n+1)}=0~~{\rm with}~~F_{2n+1}
=L_{2n+2}(N)\cdots L_4(N)\cdot L_2(N)
\end{eqnarray}
where $L_j(N)$ are linear differential operators of order $j$.

3) It was also discovered in \cite{mccoy4} by Maple calculations that
   the operators $F_n$ have in addition a direct sum decomposition
\begin{eqnarray}
&&F_{2n}=M_{2n+1}(N)\oplus \cdots \oplus M_3(N) \oplus M_1(N)\\
&&F_{2n+1}=M_{2n+2}(N)\oplus\cdots\oplus M_4(N)\oplus M_2(N)
\end{eqnarray}

4) Furthermore, the $f^{(n)}_{N,N}(t)$ have a factorization 
property first found
in \cite{mccoy4} by computer computations and proven for $n=1,2,3$ in
\cite{mccoy6}
that
\begin{eqnarray}
&& f^{(2n)}_{N,N}(t)\, \, =\,\, \,  \, 
 \sum_{m=0}^{n-1}\, K_m^{(2n)}(N)\cdot  f^{(2m)}_{N,N}(t)\, 
 +\sum_{m=0}^{2n}\,  C^{(2n)}_m(N;t) \cdot  F_N^{2n-m} \cdot  F_{N+1}^{m},   
\label{feform}\\
&&\frac{f^{(2n+1)}_{N,N}(t)}{t^{N/2}}\, \, 
=\,\,\, \sum_{m=0}^{n-1}\, K_m^{(2n+1)}(N) \cdot  
\frac{f^{(2m+1)}_{N,N}(t)}{t^{N/2}}\nonumber\\
&&\hspace{.5in}
+\sum_{m=0}^{2n+1}\, C^{(2n+1)}_m(N;t) \cdot  F_N^{2n+1-m} \cdot F_{N+1}^m,
\label{foform}
\end{eqnarray}
where  
$F_N$ is the hypergeometric function 
\begin{equation}
F_N= {}_2F_1(1/2,N+1/2;N+1;t),
\end{equation}
and $f^{(0)}_{N,N}\,\,=\,\,\,1$. 
The $K_m^{(n)}(N)$ depend only on $N$ and we note in particular that
\begin{eqnarray}
&&K^{(3)}_0(0)=\frac{1}{6}\label{K30}\\
&&K^{(4)}_0(0)=0,~~~K^{(4)}_1(0)=\frac{1}{3}\\
&&K^{(5)}_0(0)=-\frac{1}{120},~~~K^{(5)}_1(0)=\frac{1}{2}\\
&&K^{(6)}_0(0)=0,~~~K^{(6)}_1(0)=-\frac{2}{45},~~~K^{(6)}_2(0)=\frac{2}{3}
\label{K60}
\end{eqnarray}

The $\, C^{(j)}_m(N;t)$ are polynomials 
in $t$ of degree  for $N\geq 1$
\begin{eqnarray}
{\rm deg}~C^{(2n)}_m(N;t)\,\,=\,\,\,{\rm deg}~
C^{(2n+1)}_m(N;t)\,\,\,=\,\,\,\,   n \cdot (2N+1), 
\end{eqnarray} 
with $\, C^{(n)}_m(N;t)\,\,\, \sim \,\, \, \,\,t^m\,$ as $\, t\,\sim \, 0$.
which have the palindromic property
\begin{eqnarray}
\label{pale}
C^{(2n)}_m(N;t)\,\,&= \,\,\, \, \,   t^{n(2N+1)+m} \cdot \,  C^{(2n)}_m(N;1/t),
 \\
\label{palo}
C^{(2n+1)}_m(N;t)\, \,&= \,\,\,\, \,  t^{n(2N+1)+m} \cdot \,  C^{(2n+1)}_m(N;1/t).
\end{eqnarray}
Explicit formulas for the polynomials $C^{(n)}_m(N,t)$ have been
obtained in \cite{mccoy6} for $n=1,2,3$ and conjectured for $n=4$.
For example $K^{(2)}_0 =  N/2$ and
\begin{equation}
\label{initial}
C^{(2)}_m(N;t)\,\, =\,\,\, \, \,
(-1)^{m+1}  \frac{N}{2} {m \choose 2}\left[\frac{(2N+1)^2}{4N(N+1)}\right]^m   
t^m  \sum_{n=0}^{2N+1-m}c^{(2)}_{m;n}(N) t^n,
\end{equation}
where for $0\leq n\leq N-1$
\begin{eqnarray}
\label{help0}
c^{(2)}_{2;n}(N)\,\,  &=\,\, \, \, c^{(2)}_{2;\, 2N-1-n}(N)
= \sum_{k=0}^{n}\, a_k(N) a_{n-k}(N), \\
\label{dn1}
c^{(2)}_{1;n}(N)\, \,  &=\, \,  c^{(2)}_{1;\, 2N-n}(N)
 = \sum_{k=0}^{n}\, a_k(N)  a_{n-k}(N+1), 
\end{eqnarray}
and  for $  0 \leq  n  \leq   N$
\begin{equation}
c^{(2)}_{0;\, n}(N) \,\, =  \,\,\,c^{(2)}_{0;\, 2N+1-n}(N) \,\, 
=\,\,   c^{(2)}_{2;\, n}(N+1),
\label{c2n0} 
\end{equation}
and 
\begin{equation}
c_{1;N}^{(2)}(N)=\left(\frac{(1/2)_N}{N!}\right)^2\{1+2NH_N(1/2)\}
\end{equation}
where 
\begin{equation}
a_n(N)=\frac{(1/2)_N(1/2-N)_n}{(1-N)_nn!}
\end{equation}
and 
\begin{equation}
H_N(1/2)=\sum_{k=0}^{N-1}\frac{1}{k+1/2}
\end{equation}

It is certainly true (but not yet proven) that 
the factorizations (\ref{feform}) and (\ref{foform}) hold for all
$f^{(n)}_{N,N}$. The computations in \cite{mccoy6} are based on
Fuchsian differential equations for the $f^{(n)}_{N,N}(t)$. For $n=4$
the order of these equations is 20. These equations have
a direct sum decomposition into operators which are homomorphic to
symmetric powers and products of the operator which annihilates the
hypergeometric function $F_N$.

It is furthermore very suggestive that this factorization property 
has been previously seen in the correlation functions of the XXZ 
model \cite{bk}-\cite{ksts}.

The final property of the form factors to be discussed can best be
illustrated by making a ``lambda extension'', 
first introduced in \cite{mtw}, of the expansions
(\ref{ffm}) and (\ref{ffp}) by defining 
\begin{equation}
C_{-}(M,N;\lambda)
=(1-t)^{1/4}\{1+\sum_{n=1}^{\infty}\lambda^{2n}f^{(2n)}_{M,N}\}
\label{ffml}
\end{equation}
and 
\begin{equation}
C_{+}(M,N;\lambda)
=(1-t)^{1/4}\sum_{n=0}^{\infty}\lambda^{2n}f^{(2n+1)}_{M,N},
\label{ffpl}
\end{equation}
which reduce to the Ising correlations below and above $T_c$  
when $\lambda=1$. By use of a 
remarkable set of relations presented by Orrick, Nickel, Guttmann and
Perk \cite{ongp} in 2001 for small values of $M$ and $N$, 
these lambda extensions can be written in terms of theta functions \cite{ww} 
\begin{eqnarray}
&&\theta_3(u;q)=1+2\sum_{n=1}^{\infty}q^{n^2}cos 2nu\\
&&\theta_{2}(u;q)=2q^{1/4}\sum_{n=0}^{\infty}q^{n(n+1)}\cos[(2n+1)u]
=q^{1/4}e^{iu}\theta_3(u+\pi \tau/2;q)
\end{eqnarray}
and their derivatives
\begin{equation}
\frac{d}{du}\theta_n(u;q) \equiv \theta'_n(u;q)
\end{equation}
where $t^{1/2}=k$ is the modulus of elliptic functions which is related 
to the nome $q$ by
\begin{equation}
q=e^{-\pi K'(t^{1/2})/K(t^{1/2})} 
\end{equation}
and 
\begin{equation}
K(t^{1/2})=\frac{\pi}{2}{}_2F_1(1/2,1/2;t)
\end{equation}
is the complete elliptic integral of the first kind with
$K'(t^{1/2})=K((1-t)^{1/2})$ 

The simplest example given in \cite{mccoy4} is 
for the low temperature case with $M=N=0$
\begin{equation}
 C_{-}(0,0;\lambda)
=\frac{\theta_3(u;q)}{\theta_3(0;q)} \hspace{.1in}
{\rm where}\hspace{.1in} \lambda=\cos u.
\end{equation}
 For the special values $\lambda=\cos(\pi m/n)$ we
find that $C_{-}(0,0;\lambda)$ and $t$
satisfy an algebraic equation. 
Calling $C_{-}(0,0;\lambda)=\tau,$ 
it is seen in \cite{mccoy4} that for $\lambda=\cos\pi/3$
\begin{equation}
16\tau^{12}-16\tau^8-8(t-1)\tau^3+t(1-t)=0,
\end{equation}
which is a curve of genus one.
For $\lambda=\cos(\pi/4)$,
\begin{equation}
16\tau^{16}+16(t-1)\tau^8+t^2(t-1)=0
\end{equation}
is a curve of genus three which has the simple algebraic expression
\begin{equation}
C_{-}(0,0;\cos(\pi/4))=2^{-1/4}(1-t)^{1/16}[1+(1-t)^{1/2}]^{1/4}
\end{equation}

Further results in this direction are \cite{mccoy4}
\begin{eqnarray}
&&C_{-}(1,1;\cos(\pi/4))=2^{-3/4}(1-t)^{1/16}[1+(1-t)^{1/2}]^{3/4}\\
&&C_{-}(2,2;\cos(\pi/4))=2^{-5/4}(1-t)^{1/16}[1+(1-t)^{1/2}]^{5/4}
[5-(1-t)^{1/2}]/4\nonumber\\
\end{eqnarray}
Further results which follow from \cite{ongp} are given in \cite{mg}
\begin{eqnarray}
&&C_{+}(0,0;\lambda)=\frac{\theta_2(u;q)}{\theta_2(0;q)}\\
&&C_{-}(1,1;\lambda)
=-\frac{\theta'_2(u;q)}{\sin u\theta_2(0;q)\theta_3^2(0;q)}\\
&&C_{+}(1,1;\lambda)=-\frac{\theta'_3(u;q)}{\sin u\theta_3(0;q)\theta^2_2(0;q)}
\end{eqnarray}
where is  to be noted  (for $N=0,1$) that $C_{+}(N,N;\lambda)$ is
obtained from $C_{-}(N,N;\lambda)$ by the interchange
$\theta_2\leftrightarrow \theta_3$.

Many further results for various low values of $M,N$ remain (in the tradition
of Kaufman and Onsager) to be published by the authors of \cite{mccoy4}.

\section{Jimbo, Miwa and Painlev{\'e}}

The immediate object of the computation of the leading term in the
form factor expansion by Wu \cite{wu} for the row correlation 
$\langle\sigma_{0,0}\sigma_{0,N}\rangle$ and by Cheng and Wu \cite{cw}
for the general case $\langle\sigma_{0,0}\sigma_{M,N}\rangle$
was to compute the leading behavior of the correlations functions for
large separations $R=(M^2+N^2)^{1/2}$. They found that for $T<T_c$ 
the correlation decays to the limiting value (\ref{smag})
as
\begin{equation}
\langle\sigma_{0,0}\sigma_{M,N}\rangle\sim
(1-t)^{1/4}\{1-\frac{C_{-}(T)}{R^2}e^{-R/\xi_{-}(T)}\}
\label{cm}
\end{equation}
and vanishes for $T>T_c$ as
\begin{equation}
\langle\sigma_{0,0}\sigma_{M,N}\rangle
\sim(1-t)^{1/4}\frac{C_{+}(T)}{R^{1/2}}e^{-R/\xi_{+}(T)},
\label{cp}
\end{equation}
where, in addition to depending on the temperature $T$, the $R$
independent quantities $C_{\pm}(T)$ and $\xi_{\pm}(t)$ depend on the
ratio $M/N$. It is found in \cite{wu} and \cite{cw} as
$t\rightarrow 1$ that
\begin{equation}
\xi_{\pm}(T)\sim \frac{A_{\xi,\pm}}{1-t}
\end{equation}
\begin{equation}
C_{-}(T)\sim \frac {A_{-}}{(1-t)^2}
\end{equation}
and 
\begin{equation}
C_{+}(T)\sim\frac{A_{+}}{(1-t)^{1/2}},
\end{equation}
where again the amplitudes $A_{\xi.\pm}$ and $A_{\pm}$ depend on the
ratio $M/N$. Neither of these asymptotic leading terms reduces to the
result valid for $T=T_c$ (i.e. $t=1$) where in \cite{wu} Wu 
found that the diagonal correlation has the
leading behavior for large $N$
\begin{equation}
\langle\sigma_{0,0}\sigma_{N,N}\rangle\sim \frac{A_{T_c}}{N^{1/4}}
\label{ctc}
\end{equation}
and
\begin{equation}
A_{T_c}=2^{1/12}e^{3\zeta'(-1)}
\label{atc}
\end{equation}
with $\zeta'(-1)$ the derivative of Riemann's zeta function at $-1$.

The history of the result (\ref{ctc})  is romantic in its own way. In
the original 1949 paper of \cite{ko} there is a remark that the
diagonal correlation vanishes ``slowly''. In 1959 Fisher \cite{fisher}
derived the exponent $1/4$ and remarked in footnote 8 that 

\vspace{.1in}
Onsager, private communication, has derived exact expressions for the
correlations along the main diagonal $\cdots$ 

\vspace{.1in}

\noindent This computation was never published and perhaps there is
another typescript out there waiting to be discovered. 
 
Wu \cite{wu} also found the large $N$ behavior of the row correlation
$\langle\sigma_{0,0}\sigma_{0,N}\rangle$, which has the same 
dependence on $N$ as (\ref{ctc}) but with an amplitude 
\begin{equation}
A_{\rm{row}}=A_{T_c}(\cosh 2E^h/k_BT_c)^{1/4}.
\end{equation}

The first purpose of the paper \cite{mccoy3} was to connect
the three different asymptotic behaviors (\ref{cm}), (\ref{cp}) and
(\ref{ctc}) by defining an interpolating function, traditionally
called a scaling function,
\begin{equation}
G_{\pm}(r)=\lim_{M,N\rightarrow \infty,t\rightarrow 1}(1-t)^{-1/4}
\langle\sigma_{0,0}\sigma_{M,N}\rangle
\end{equation}
with 
\begin{equation}
\left[\left(\frac{\sinh 2 E^h/k_BT_c}{\sinh 2 E^v/k_BT_c}\right)^{1/2}M^2
+\left(\frac{\sinh 2 E^v/k_BT_c}{\sinh 2
  E^h/k_BT_c}\right)^{1/2}N^2\right]^{1/2}(1-t)=r
\hspace{.1in} {\rm fixed}.
\label{rdef}
\end{equation}
For this purpose the exponential representation of the correlation
functions was
derived. When the scaling function was computed it was discovered
that $G_{\pm}(r)$ is expressed in terms of a Painlev{\'e} equation 
of the third kind 
\begin{equation}
\frac{d^2\eta}{d\theta^2}=\frac{1}{\eta}\left(\frac{d\eta}{d\theta}\right)^2
-\frac{1}{\theta}\frac{d\eta}{d\theta}+\eta^3-\eta^{-1}
\end{equation}
as
\begin{equation}
G_{\pm}(r)=\frac{1\mp \eta(r/2)}{2\eta(r/2)^{1/2}}
\exp\frac{1}{4}\int_{r/2}^{\infty}d\theta \theta
\frac{(1-\eta^2)^2-(\eta')^2]}{\eta},
\end{equation}
with the boundary condition
\begin{equation}
\eta(\theta)\sim 1-\frac{2}{\pi}\lambda K_0(2\theta)
~~{\rm as}~~\theta\rightarrow \infty,
\end{equation}
where $K_0(2\theta)$ is the modified Bessel function and $\lambda=1$.

This result was first announced in \cite{mccoy1} and \cite{mccoy2}.Two
different proofs were given. The first, in
\cite{mccoy3}, is based on Myers' work \cite{myers} on the scattering of
electromagnetic radiation from a strip  and the second \cite{mtw} is
based on a direct manipulation of the exponential representation
in the scaling limit.

It is at this point that I first learned of the existence of Sato,
Miwa and Jimbo when in 1977 I received in the mail (how long ago it
was that papers were sent by mail) a letter by the three of them with title
``Studies on holonomic quantum fields II'' \cite{smj2} which
generalized several of the results of \cite{mccoy3}  and made clear the
relation of the Painlev{\'e} III equation with the massive Dirac
equation. This letter was followed by many more where the only change
in the title was that the Roman numeral was different and by a 
series of 5 papers with the title ``Holonomic quantum
field theory''\cite{smj}. These papers culminated in the groundbreaking
paper ``Studies on holonomic quantum fields XVII'' \cite{jm}
where it is derived that the diagonal Ising correlation function for a
general temperature on the lattice and not in the scaling limit
satisfies the sigma form of the Painlev{\'e} VI equation
\begin{eqnarray}
&&\Bigl( t\, (t-1) \frac{d^2\sigma}{dt^2} \Bigr)^2\, =\,\, \nonumber  \\
&&\quad N^2 \Bigl( (t-1) \frac{d\sigma}{dt} -\sigma \Bigr)^2-
4\frac{d\sigma}{dt}
\Bigl((t-1)\frac{d\sigma}{dt}-\sigma -1/4   \Bigr)
\Bigl(t\frac{d\sigma}{dt}-\sigma   \Bigr). 
\label{pvi}
\end{eqnarray}
The diagonal correlation is related to  $\sigma$ for $T >T_c$ by
\begin{eqnarray}
\label{sigma-bas}
\sigma(t)=t(t-1) \cdot {\frac{d}{dt}}\log\langle\sigma_{0,0}\sigma_{N,N}\rangle
-1/4,
\end{eqnarray}
with the boundary condition at $t=0$ of
\begin{equation}
\langle\sigma_{0,0}\sigma_{N,N}\rangle=t^{N/2}\frac{(1/2)_N}{N!}+O(t^{1+N/2}),
\end{equation}
and for $T <T_c$ by
\begin{eqnarray}
\label{sigma-haut}
\sigma(t)\,=\,\, t(t-1)  \cdot{\frac{d}{dt}}\log\langle
\sigma_{0,0}\sigma_{N,N}\rangle-t/4
\end{eqnarray}
with the boundary condition
\begin{equation}
\langle\sigma_{0,0}\sigma_{N,N}\rangle
=(1-t)^{1/4}\{1-\frac{t^{N+1}}{2N+1}\left(\frac{(1/2)_{N+1}}{(N+1)!}\right)^2
+O(t^{N+2})\}
\end{equation}
where $(a)_N=a(a+1)\cdots(a+N-1)$ for $1\leq N$ and $(a)_0=1$ is
Pochammer's symbol.
These boundary conditions are obtained from the leading terms of
(\ref{fdm}) and (\ref{fdp}) as $t\rightarrow 0$.
Furthermore the lambda extensions (\ref{ffml}) and (\ref{ffpl}) satisfy
the same Painlev{\'e} VI equation (\ref{pvi}) where the $\lambda$
appears as a boundary condition.

The six Painlev{\'e} equations have a long history
\cite{ince},\cite{gar1}. They are defined
as those second-order nonlinear equations the location of 
whose branch points and essential singularities (but not poles) are
independent of the boundary conditions and which cannot be reduced to
simpler functions. Painlev{\'e} obtained three of these equations
\cite{pain} and Gambier \cite{gamb} obtained the remaining three including
the PVI equation which in the general case has four parameters.
However, the specific case of Painlev{\'e} VI
needed for the Ising model (\ref{pvi}) had already been obtained by
Picard \cite{pic} in 1889. Subsequent to the discovery that this PVI
equation characterizes the diagonal Ising model, this equation has
appeared in many contexts \cite{hit}-\cite{mazz} ranging from 
Poncelet polygons to mirror symmetry. The sigma form of the Painlev{\'e} 
equations was first obtained by Okamoto \cite{oka}.



\section{The susceptibility}

The second purpose of the paper \cite{mccoy3} was to begin the study of
the magnetic susceptibility at zero magnetic field $\chi(T)$, which is
computed in terms of the correlation functions as
\begin{equation}
k_BT\chi(T)=\sum_{M=-\infty}^{\infty}\sum_{N=-\infty}^{\infty}
\{\langle\sigma_{0,0}\sigma_{M,N}\rangle-{\mathcal M}^2\},
\label{susp}
\end{equation}
where ${\mathcal M}^2$ is the square of the spontaneous magnetization
which was given in (\ref{smag}). In order to evaluate the sums in
(\ref{susp}) the exponential forms (\ref{eformm}) and (\ref{eformp})
which were the basis of computing the Painlev{\'e} III equation cannot
be used and instead the exponentials must be expanded into the form
factor representations (\ref{ffm}) and (\ref{ffp}). Using these forms
the sums over $M$ and $N$ are easily evaluated  as geometric series and the
susceptibility is written as the infinite sum of $n$ ``particle''
contributions
\begin{eqnarray}
k_BT\chi_+(T)&=&(1-t)^{1/4}t^{-1/4}\sum_{j=0}^{\infty}{\chi}^{(2j+1)}(T)
~~{\rm for}~ T>T_c\label{chip}\\ 
k_BT\chi_-(T)&=&(1-t)^{1/4}\sum_{j=1}^{\infty}{\chi}^{(2j)}(T)
~~{\rm for}~T<T_c.\label{chim}
\end{eqnarray}

In \cite{mccoy3} the terms $\chi^{(n)}(T)$ for $n=1,2,3,4$ were
studied. In the scaling limit the scaled $\chi^{(n)}(T)$ for
general $n$ were given by Nappi \cite{nap} in 1978. For 
arbitrary temperature the results in the isotropic case 
were obtained by Nickel \cite{ni1} and \cite{ni2} and for $E^v\neq E^h$
in \cite{ongp}
\begin{equation}
{ \chi}^{(j)}(T)={\cot^j \alpha\over j!}
\int_{-\pi}^{\pi}{d\omega_1\over 2\pi}\cdots 
\int_{-\pi}^{\pi}{d\omega_{j-1}\over 2\pi}
\left(\prod_{n=1}^j{1\over \sinh \gamma_n}\right)
H^{(j)}
{1+\prod_{n=1}^jx_n\over 1-\prod_{n=1}^j x_n},
\label{bernie}
\end{equation}
with
\begin{equation}
x_n=\cot^2 \alpha \left[\xi -\cos \omega_n-
\sqrt {(\xi-\cos \omega_n)^2-(\cot \alpha)^{-4}}\right]
\label{10xn}
\end{equation}
\begin{equation}
\sinh \gamma_n=\cot^2 \alpha\sqrt{(\xi-\cos \omega_n)^2-(\cot \alpha)^{-4}},
\end{equation}
where
\begin{equation}
\cot \alpha=\sqrt{ s_h/s_v}
\label{10calpha}
\end{equation}
\begin{equation}
\xi=(1+s_h^{-2})^{1/2}(1+s_v^2)^{1/2}
\label{10xid}
\end{equation}
\begin{equation}
s_v=\sinh 2E^v/k_BT~~~~s_h=\sinh 2E^h/k_BT
\end{equation}

\begin{equation}
H^{(j)}=
\left(\prod_{1\leq i<k\leq j} h_{ik}\right)^2
\end{equation}
with
\begin{equation}
h_{ik}=\cot\alpha {\sin{1\over 2}(\omega_i-\omega_k)\over 
\sinh{1\over 2}(\gamma_i-\gamma_k)}
={1\over \cot\alpha}{\sinh {1\over 2}(\gamma_i-\gamma_k)\over
  \sin{1\over 2}(\omega_i+\omega_k)},
\end{equation}
and $\omega_j$ is defined in terms of the remaining $\omega_i$ from
$\omega_1+\cdots \omega_j=0~{\rm mod}~2\pi.$
We note in particular that for $E^v=E^h$ 
\begin{equation}
\chi^{(1)}(t)=\frac{t^{1/4}}{(1-t^{1/4})^2}
\label{chi1}
\end{equation}
with $t$ given by (\ref{tp}) and
\begin{equation}
\chi^{(2)}(t)=\frac{(1+t)E(t^{1/2})-(1-t)K(t^{1/2})}{3\pi(1-t^{1/2})(1-t)}
\label{chi2}
\end{equation}
with $t$ given by (\ref{tm}).

\subsection{The amplitude of the susceptibility divergence}

The study of the susceptibility from the form factor expansions was
initiated in 1973 in \cite{mccoy1} where it was demonstrated 
that as $T\rightarrow T_c\pm$  the
susceptibility diverges as
\begin{equation}
k_BT\chi(T)_{\pm}\sim C_{\pm}\left| 
\frac{s^{-1}-s}{2}\right| ^{-7/4}{\sqrt{2}}.
\end{equation}
where in the isotropic case 
\begin{equation}
s=\sinh 2E/k_BT.\label{sdef}
\end{equation}  
The constants 
$C_{-}$ and $C_{+}$ are different and are given as infinite series
\begin{equation}
C_{-}=\sum_{n=1}^{\infty}C^{(2n)} \hspace{.5in}
C_{+}=\sum_{n=0}^{\infty}C^{(2n+1)}
\label{csum}
\end{equation}
where the $C^{(n)}$ are $n-$fold integrals 
coming from the form factor expansion and have been
studied both numerically for  $n=1,\cdots, 5$
\cite{mccoy1},\cite{mccoy3}.
The first term in each of (\ref{csum})  has been analytically evaluated in 
\cite{mccoy1},\cite{mccoy3}
\begin{equation}
C^{(1)}=1\hspace{.5in} C^{(2)}=\frac{1}{12\pi}
\end{equation}
and the next leading term was evaluated by Tracy
 \cite{tracy1}  
as
\begin{equation}
C^{(3)}=\frac{1}{2\pi^2}\left(\frac{\pi^2}{3}+2-3{\sqrt 3}{\rm
  Cl}_2(\pi/3)\right)
\end{equation}
where 
\begin{equation}
{\rm Cl}_2(\theta)=\sum_{n=1}^{\infty}\frac{\sin n \theta}{n^2}
\end{equation}
is Clausen's function and
\begin{equation}
C^{(4)}=\frac{1}{16\pi^3}\left(\frac{4\pi^2}{9}-\frac{1}{6}
-\frac{7}{2}\zeta(3)\right).
\end{equation}
In the tradition of Onsager
and Kaufman \cite{baxter} the details are only in an unpublished typescript.
A curious feature of these results is that the ratio $C_{+}/C_{-}$ is
found to be closely approximated by $12\pi$ and the second terms are
approximately three orders of magnitude less than the leading term.
The study of the  constants $C_{-}$ and $C_{+}$ has been 
continued by high precision numerical computations \cite{ongp} 
and the most recent evaluation \cite{cgnp} in 2011 is to  
an incredible 104 places. This is one of the most precisely determined
constants in all of mathematical physics.

However, the $\chi^{(n)}(w)$ have singularities at other points
besides $\sinh E/k_BT=\pm 1$ and the determination of the analytic 
properties of the magnetic susceptibility as a function of temperature
has become  the most challenging problem is the field.

\subsection{Nickel singularities and the natural boundary conjecture}

The first studies of analytic properties after the initial
computations of \cite{mccoy3} were made in 1999 \cite{ni1}
and 2000 \cite{ni2} when Nickel demonstrated for the isotropic case 
$E^v=E^h=E$ that the integrals (\ref{bernie})   $\chi^{(n)}$ have
singularities in the complex $T$ plane on the curve
\begin{equation}
|\sinh 2E/k_BT|=1,
\end{equation}
which is the same curve on which the four Pfaffians of Kaufman's original
evaluation \cite{kauf} of the Ising partition function vanish. This
was extended to the general case $E^v\neq E^h$ in \cite{ongp} where
the singularities of $\chi^{(n)}(T)$ are at
\begin{eqnarray}
&&\cosh 2E^v/k_BT\cosh 2E^h/k_BT\nonumber\\
&&-\sinh 2E^h/k_BT \cos (2\pi j/n)
-\sinh 2E^v/k_BT \cos (2\pi k/n)=0.
\label{location}
\end{eqnarray}
with 
\begin{equation}
0\leq j,k \leq [n/2],~~~j=k=0~~ {\rm excluded} 
\end{equation}
where $[x]$ is the integer part of $x$ and for $n$ even $j+k=n/2$ is
also excluded.
In terms of the variable used in \cite{jm1}-\cite{jm8} for the
isotropic lattice with $s$ given by (\ref{sdef})
\begin{equation}
w^{-1}=2(s+s^{-1})
\end{equation}
these singularities  for $n=3,4,5,6$ are given in table \ref{tab1}

\begin{table}[h!]
\label{tab1}
\caption{The Nickel singularities of $\chi^{(n)}$ for $n=3,4,5,6$}

\begin{center}
\begin{tabular}{|l|l|}\hline
n&w\\ \hline
3&$-1/2,~1$\\
4&$\pm 1/2$\\
5&$-1,~\frac{-1\pm \sqrt{5}}{4},~\frac{3\pm \sqrt{5}}{2}$\\
6&$\pm 1,~\pm 1/3$\\ \hline
\end{tabular}

\end{center}
\end{table}
\noindent where we note that $\sinh 2E/k_BT$ is real for $-1/4\leq w
\leq 1/4$ and is complex with $|\sinh 2E/k_BT|=1$ for $1/4<|w|$. 
If we call $\epsilon$ the deviation from the singular temperatures
$T^{(j)}_{m,m'}$ determined by (\ref{location}), then for $T>T_c$ 
the singularity in
${\chi}^{(2j+1)}(T)$ is 
\begin{equation}
\epsilon^{2j(j+1)-1}\ln \epsilon
\label{singodd}
\end{equation}  
and for $T<T_c$ the singularity in  ${\chi}^{(2j)}(T)$ is 
\begin{equation}
\epsilon^{2j^2-3/2}.
\label{singeven}
\end{equation}
 
It is striking that the number of singularities increases with
$n$ and becomes dense in the limit $n\rightarrow \infty$. This 
feature led Nickel to the conclusion that unless cancellations
occur there will be a natural boundary in the susceptibility $\chi(T)$
in the complex $T$ plane at the location (\ref{location}).  
The existence of a natural boundary in the complex
temperature plane is not contemplated in the scaling theory of
critical phenomena. 

\subsection{Fuchsian equations}

The next step in the study of the susceptibility was begun in 2005
 \cite{jm1} and has continued in the series of papers 
\cite{jm2}-\cite{jm8}.
In these papers exact Fuchsian differential equations for the $\chi^{(n)}(T)$
in the isotropic case $E^v=E^h$ are determined by use of Maple by
first expanding the integrals in an appropriate variable such as
$w$ or $w^2$ and then using Maple programs which obtain ODE's
from these series. The resulting differential equations have
very special properties such as being globally nilpotent \cite{jm7} which allow
for extensive analysis to be carried out. These studies have uncovered 
several new and important features of the susceptibility; namely that
the $\chi^{(n)}(w)$ have a direct sum decomposition and that they have
 further singularities beyond those of (\ref{location}).

\subsubsection{Direct sum decompositions}

In \cite{jm4} and \cite{jm5} it is shown for $1\leq n \leq 6$ that
$\chi^{(n)}(w)$ have the same direct sum
decomposition seen already in the diagonal form factors
\begin{eqnarray}
&&\chi^{(2n)}(w)=\sum_{m=1}^{n-1}K^{(2n)}_m\chi^{(2m)}(w)+\Omega^{(2n)}(w)
\label{chievendsum}\\
&&\chi^{(2n+1)}(w)
=\sum_{m=1}^{n-1}K^{(2n+1)}_m\chi^{(2m+1)}(w)+\Omega^{(2n+1)}(w)
\label{chiodddsum}
\end{eqnarray}
where the $\Omega^{(n)}(w)$ satisfy Fuchsian equations of order $m$
\begin{equation}
L^{(n)}_m \cdot \Omega^{(n)}=0
\label{lnm}
\end{equation}
with
\begin{eqnarray}
\begin{array}{ccccc}
n&3&4&5&6\\
m&6&8&29&46
\end{array}
\end{eqnarray}
The $K^{(n)}_j$ are constants which for $n=3,4,5,6$ coincide with the
values of $K^{(n)}_m(0)$ given in (\ref{K30})-(\ref{K60}).

The operators in (\ref{lnm}) factorize further. For
$L^{(3)}_6$ and $L^{(4)}_8$ we have 
we have
\begin{equation}
L^{(3)}_6=L^{(3)}_3\cdot L^{(3)}_2 \cdot L^{(3)}_1
\end{equation}
and
\begin{equation}
L^{(4)}_8=L^{(4)}_4 \cdot L^{(4)}_1\cdot \left(L^{(4)}_{1;a}\oplus
L^{(4)}_{1;b} \oplus L^{(4)}_{1;c} \right)
\end{equation}
where the numeral in the subscript indicates the order of the
operators which are given in \cite{jm3} and \cite{jm4}.
The operator 
$L^{(5)}_{29}$ has been found in \cite{exp},\cite{jm9} and \cite{jm8}
to have the factorization
\begin{equation}
L^{(5)}_{29}=L^{(5)}_5\cdot L^{(5)}_{12} \cdot L^{(5)}_1\cdot L^{(5)}_{11}
\end{equation}
where $L^{(5)}_{11}$ has the further direct sum decomposition (A.1) of
\cite{jm8}
\begin{equation}
L^{(5)}_{11}=(Z_2 \cdot N_1)\oplus V_2\oplus (F_3\cdot F_2\cdot L^s_1)
\end{equation} 
Similarly in
(56) and (57) of \cite{jm5} the operator $L^{(6)}_{46}$ is shown to
have the decomposition
\begin{equation}
L^{(6)}_{46}=L^{(6)}_6\cdot L^{(6)}_{23} \cdot L^{(6)}_{17}
\end{equation}
where $L^{(6)}_{17}$ has a direct sum decomposition into the sum of
four operators but the possible reducibility of $L^{(6)}_{23}$ has not
yet been determined due to computational complexity.

\subsubsection{Singularities}

The location of the singularities of the operators $L^{(n)}_m$ are
obtained by examining the roots of the polynomial multiplying the
highest derivative $d^m/dw^m$ and this analysis shows that there are
further singularities beyond the singularities at $w=\pm 1/4, \infty$ and
the Nickel singularities (\ref{location}).

In \cite{jm2} that
the differential equation for $\chi^{(3)}(w)$ admits  additional  
singularities at
\begin{equation}
w=\frac{-3\pm i\sqrt{7}}{8}
\end{equation}
which correspond to
\begin{eqnarray}
&&s=\frac{-1\pm i\sqrt{7}}{4},~~~|s|=\frac{1}{\sqrt 2}\label{inside}\\
&&s=\frac{-1\pm i\sqrt{7}}{2},~~~|s|=\sqrt{2}
\end{eqnarray}
where we note the the singularity at (\ref{inside}) is inside the unit
circle $|s|$=1 and thus cannot appear in the principle sheet of the
integral for $\chi^{(3)}$ which is analytic for $|s|<1$.

There are no additional singularities in $\chi^{(4)}(w)$ and the 
singularities of $\chi^{(5)}(w)$ are shown in (34) of  \cite{exp} to be 
at the roots of following polynomial
\begin{eqnarray}
&&w^{33}(1-4w)^{22}(1+4w)^{16}(1-w)^4(1+2w)^4(1+3w+4w^2)\nonumber\\
&&(1+w)(1-3w+w^2)(1+2w-4w^2)\nonumber\\
&&(1-w-3w^2+4w^3)(1+8w+20w^2+15w^3+4w^4)(1-7w+5w^2-4w^3)\nonumber\\
&&(1+4w+8w^2)(1-2w)
\label{chi5roots}
\end{eqnarray}
The singularities located by the roots of the first line in
(\ref{chi5roots}) are identical with the location of singularities
of $\chi^{(3)}$ and the roots of the second line are the Nickel
singularities of of $\chi^{(5)}$. Most of the remaining singularities 
correspond to complex values of $s$ not on $|s|=1$.



\section{Diagonal susceptibility}

The integrals (\ref{bernie}) for the $n$ particle contribution to 
the susceptibility $\chi^{(n)}(T)$ are quite complex and the
Maple-based studies cannot be extended much beyond their present
limits. Therefore it would be of great utility if a simpler set of
integrals could be found which would still incorporate all
significant analytic features of the the $\chi^{(n)}$. Several 
such simplified modifications of the integrals have been studied
\cite{jm6}
but by far the most natural case is to restrict the two dimensional 
sum over the lattice positions $M,N$ in (\ref{susp}) to the lattice
diagonal $M=N$ and thus to consider the susceptibility that will
result if a magnetic field is applied only to the diagonal
\begin{equation}
k_BT\chi_d(t)=\sum_{N=-\infty}^{\infty}\{\langle
\sigma_{0,0}\sigma_{N,N}\rangle-{\mathcal M}^2\},
\end{equation}
where the dependence on $T$ is now for all $E^v$ and $E^h$ in terms of 
the  single variable $t$  defined by (\ref{tm}) for $T<T_c$ and by
(\ref{tp}) for $T>T_c$. 

This diagonal susceptibility has been studied in \cite{mccoy9} and
\cite{mccoy11} and has been found to have the remarkable
simplification over the bulk susceptibility that all singularities of
the differential equations are at $s=0,\infty$ and $|s|=1$. There are
no other complex singularities for $|s|\neq 1$ such as appear in
$\chi^{(n)}(t)$. Furthermore $\chi^{(3)}_d(t)$ and $\chi^{(4)}_d(t)$ have been
found to be explicitly expressed in terms of generalized
hypergeometric functions ${}_{p+1}F_p$.

\subsection{Integral representations}
From the integral expressions for
$f^{(n)}_{N,N}(t)$ of (\ref{fdm}) and (\ref{fdp}) given in \cite{mccoy5}
and \cite{mccoy4},
we find in \cite{mccoy9} the
expansion for $T<T_c$
\begin{equation}
k_BT\chi_{d,-}(t)=(1-t)^{1/4}\sum_{n=1}^{\infty}\chi_d^{(2n)}(t)
\label{chidm}
\end{equation}
and for $T>T_c$
\begin{equation}
k_BT\chi_{d,+}(t)=(1-t)^{1/4}\sum_{n=0}^{\infty}\chi_d^{(2n+1)}(t),
\label{chidp}
\end{equation}
where
\begin{eqnarray}
&&{\chi}^{(2n)}_{d}(t)= 
 {{  t^{n^2}} \over {  (n!)^2 }}  
{{1 } \over {\pi^{2n} }} \cdot  
\int_0^1 \cdots  \,\int_0^1\prod_{k=1}^{2n}\,  dx_k  
\cdot   {1\, +t^n\, x_1\cdots x_{2n}\over
  1\,-t^n\, x_1 \cdots x_{2n}}\nonumber \\ 
&&\times
\prod_{j=1}^n\left({x_{2j-1}(1-x_{2j})(1-tx_{2j})\over 
x_{2j}(1-x_{2j-1})(1\, -t\, x_{2j-1})}\right)^{1/2}
\prod_{1 \leq j \leq n}
\prod_{1 \leq k \leq n}(1\, -t\, x_{2j-1}\, x_{2k})^{-2}\nonumber\\
&&\times\prod_{1 \leq j<k\leq n}(x_{2j-1}-x_{2k-1})^2\, (x_{2j}-x_{2k})^2
\label{echidm}
\end{eqnarray}
and for $T>T_c$
\begin{eqnarray}
&&{\chi}^{(2n+1)}_{d}(t)={t^{n(n+1))}
 \over \pi^{2n+1} n!  (n+1)! } \cdot 
\int_0^1 \cdots \int_0^1 \, \, \prod_{k=1}^{2n+1}  dx_k\nonumber\\
&& \times {1\, +t^{n+1/2}\,x_1\cdots x_{2n+1}\over
 1\, -t^{n+1/2}\, x_1\cdots x_{2n+1}} \cdot 
 \prod_{j=1}^{n}\, \Bigl((1-x_{2j})(1\,-t\,x_{2j})\cdot
 x_{2j}\Bigr)^{1/2} \nonumber \\
&&\times\prod_{j=1}^{n+1} \, 
\Bigl((1\,  -x_{2j-1})(1\,-t\, x_{2j-1}) \cdot x_{2j-1}\Bigr)^{-1/2} 
\nonumber\\
&&\times
\prod_{1\leq j\leq n+1}\prod_{1\leq k \leq n}\, 
(1\, -t\, x_{2j-1}\, x_{2k})^{-2}\nonumber \\
&&\times\prod_{1 \leq j<k\leq n+1}(x_{2j-1} -x_{2k-1})^2
\prod_{1\leq j<k\leq n}(x_{2j}-x_{2k})^2. 
\label{echidp}
\end{eqnarray} 
The expressions (\ref{echidm}) and (\ref{echidp}) are, indeed,
much simpler than the corresponding expressions for
${\chi}^{(n)}$ given in (\ref{bernie}). In particular
\begin{equation}
\chi^{(1)}_d(t)=\frac{1}{1-t^{1/2}}
\end{equation}
and
\begin{equation}
\chi^{(2)}_d(t)=\frac{t}{4(1-t)}
\label{chi2d}
\end{equation}
which are simpler than (\ref{chi1}) and (\ref{chi2})
respectively. 
Most noticeable is that $\chi^{(2)}(w)$ in (\ref{chi2}) has
a logarithmic singularity at $t=1~(w=1/4)$ while $\chi^{(2)}_d(t)$ 
in (\ref{chi2d}) does not.

\subsection{Root of unity singularities}

In addition to the singularity at $t=1$ it is straightforward to see 
from the integral expressions (\ref{echidm}) and (\ref{echidp}) that
$\chi_d^{(2n)}(t)$ has singularities at 
\begin{equation}
 t_0^n=1 \label{t0m}
\end{equation}
of the form
\begin{equation}
\epsilon^{2n^2-1}\ln \epsilon
\end{equation} 
and $\chi_d^{(2n+1)}(t)$ has singularities 
\begin{equation}
t_0^{n+1/2}=1
\label{t0p}
\end{equation}
of the form
\begin{equation}
\epsilon^{(n+1)^2-1/2}
\end{equation}
where $\epsilon$ is the deviation from $t_0$.
These are the analogues for the diagonal susceptibility of the Nickel
singularities of the bulk susceptibility $\chi^{(n)}$ of (\ref{location}).

\subsection{Direct sum decomposition}

The $\chi_d^{(n)}(t)$ have the same direct sum
decomposition seen already in the diagonal form factors and 
$\chi^{(n)}(w)$
\begin{eqnarray}
&&\chi_d^{(2n)}(t)
=\sum_{j=1}^{n-1}K^{(2n)}_{d;j}\chi_d^{(2j)}(t)+\Omega_d^{(2n)}(t)
\label{chidedsum}\\
&&\chi_d^{(2n+1)}(t)=\sum_{j=1}^{n-1}K^{(2n+1)}_{d;j}\chi_d^{(2j+1)}(t)
+\Omega_d^{(2n+1)}(t)\label{chidodsum}
\end{eqnarray}
where $K_{d;j}^{(n)}$ are constants.
However, unlike $\chi^{(n)}(w)$, the operators $L^{(n)}_d$ which annihilate
$\Omega_d^{(n)}(t)$ have a further direct sum decomposition
\begin{eqnarray}
L^{(3)}_{d;5}=L^{(3)}_{d;2}+L^{(3)}_{d;3}\hspace{.3in}{\rm and}\hspace{.3in}
L^{(4)}_{d;7}=L^{(4)}_{d;3}+L^{(4)}_{d;4}
\end{eqnarray}

\subsection{Results for $\chi^{(3)}_d(t)$}

For $\chi^{(3)}_d(t)$ we explicitly find by 
combining \cite{jm7} and \cite{mccoy9} 
and setting $x=t^{1/2}$ that
\begin{equation}
\chi^{(3)}_d(x)=\frac{1}{3}\chi^{(3)}_{d;1}(x)+\frac{1}{2}\chi^{(3)}_{d;2}(x)
-\frac{1}{6}\chi^{(3)}_{d;3}(x)
\label{chi3sum}
\end{equation}
where
\begin{equation}
\chi^{(3)}_{d;1}(x)=\frac{1}{1-x}=\chi_d^{(1)}(x)
\label{chi31}
\end{equation} 
\begin{equation}
\chi^{(3)}_{d;2}(x)=\frac{1}{(1-x)^2}{}_2F_1(1/2,-1/2;1;x^2)
-\frac{1}{1-x}{}_2F_1(1/2,1/2;1;x^2)
\label{chi32}
\end{equation}
and
\begin{eqnarray}
\chi^{(3)}_{d;3}(x)=&&\frac{(1+2x)(x+2)}{(1-x)(x^2+x+1)}\left[F(1/6,1/3;1;Q)^2
\right.\nonumber\\
&&\left.+\frac{2Q}{9}F(1/6,1/3;1;Q)(F(7/6,4/3;2;Q)\right]
\label{chi33}
\end{eqnarray}
with
\begin{equation}
Q=\frac{27}{4}\frac{(1+x)^2x^2}{(x^2+x+1)^3}
\label{qdef}
\end{equation}
where we note that
\begin{equation}
1-Q=\frac{(1-x)^2(1+2x)^2(2+x)^2}{4(1+x+x^2)^3}
\label{1mq}
\end{equation}
From (\ref{chi3sum}) and (\ref{chi31}) we see that 
in (\ref{chidodsum}) we have $K^{(1)}_1=1/3$.

We see from (\ref{echidp}) that $\chi^{(3)}_{d}(x)$ vanishes when
$x\rightarrow 0$ as $x^4$. However, $\chi^{(3)}_{d;1}(x)$ and 
$\chi^{(3)}_{d;3}(x)$ are constant as $x\rightarrow 0$ and 
$\chi^{(3)}_{d;2}(x)$ vanishes linearly in $x$. The three constants in
(\ref{chi3sum}) are determined by matching with the $x^4$ behavior of
$\chi^{(3)}_{d}$ and this requires that the three constants will solve
a set of five (overdetermined) linear equations.

As $x\rightarrow 1$ we find that $\chi^{(3)}_d$ diverges as
\begin{equation}
\chi^{(3)}_d(x)=\frac{1}{1-x}\left(\frac{1}{3}
-\frac{5\pi}{18\Gamma^2(5/6)\Gamma^2(2/3)}+
\frac{4\pi}{\Gamma^2(1/6)\Gamma^2(1/3)}\right)=\frac{0.016329\cdots}{1-x}
\end{equation}

Furthermore $\chi^{(3)}_{d;3}(x)$ has an additional singularity at
$x\rightarrow e^{\pm 2 \pi i/3}$ which, to leading order is 
\begin{equation}
\chi^{(3)}_{d;3{\rm sing}}\rightarrow 
\frac{3^{4/5}16}{35 \pi}e^{\pm 5 \pi i/12}\epsilon^{7/2}
\end{equation}

\subsection{Results for $\chi^{(4)}_d(t)$}

These results have been extended in \cite{mccoy9} and \cite{mccoy11} 
to $\chi_{d}^{(4)}(t)$ where is is shown that
\begin{equation}
\chi^{(4)}_d(t)=\frac{1}{2^3}\chi^{(4)}_{d;1}(t)
+\frac{1}{3\cdot 2^3}\chi^{(4)}_{d;2}(t)-\frac{1}{2^3}\chi^{(4)}_{d;3}(t)
\end{equation}
where
\begin{equation}
\chi^{(4)}_{d;1}(t)=\chi^{(2)}_d(t)
\end{equation}
\begin{eqnarray}
\chi^{(4)}_{d;2}(t)=&& 
\frac{1+t}{(1-t)^2}{}_2F_1(1/2,-1/2;1;t)^2-{}_2F_1(1/2,1/2;1,t)^2
\nonumber\\
&&-\frac{2t}{1-t}F(1/2,1/2;1;t){}_2F_1(1/2,-1/2;1;t)
\end{eqnarray} 
and 
\begin{equation}
\chi^{(4)}_{d;4}(t)=A_3\cdot {}_4F_3([1/2,1/2,1/2,1/2];[1,1,1]t^2)
 \end{equation}
with
\begin{eqnarray}
A_3&=&2(1+t)t^3D^3_t+\frac{2}{3}\frac{16t^2-t-11}{t-1}t^2d^2_t\nonumber\\
&&+\frac{1}{3}\frac{31t^2-4t-11}{t-1}tD_t+t
\end{eqnarray}

The singular behavior as $t\rightarrow 1$ of $\chi^{(4)}_{d;1}(t)$ 
and $\chi^{(4)}_{d;3}$ is easily obtained and the singularity of
$\chi^{(4)}_{d;4}$ at $t=1$ is obtained by use of the analytic
continuation formula of B{\"u}hring \cite{buhring}. The final result 
\cite{mccoy11} is
that as $t\rightarrow 1$
\begin{eqnarray}
\chi^{(4)}(t)&\rightarrow&
\frac{1}{8(1-t)}\left(1-\frac{1}{3\pi^2}[64+16(3I_1-4I_2)]\right)\nonumber\\
&&+\frac{7}{16\pi^2}\ln
      \frac{16}{1-t}-\frac{1}{16\pi^2}\ln^2\frac{16}{1-t}
\end{eqnarray}
where
\begin{equation}
3I_1-4I_2=-2.2128121\cdots
\end{equation}
has been given to 100 digits.

At the root of unit singularity $t=-1$ the leading singular behavior is
\begin{equation}
\chi^{(4)}_d\rightarrow 
\frac{1}{26880}(1+t)^7\ln(1+t)
\end{equation}

\subsection{$\chi^{(5)}_d(t)$}

The ODE satisfied by $\chi^{(5)}_d(x)$ has been studied in
\cite{mccoy11} modulo a large prime. It is found that the minimal order
ODE is of order 19 and that the operator $L^{(5)}_{d;19}$ has the
decomposition
\begin{equation}
L^{(5)}_{d;19}=L^{(3)}_{d;2}\oplus L^{(5)}_{d;17}
\end{equation}
where $L^{(3)}_{d;2}$ is the second order operator which annihilates 
$\chi^{(3)}_{d;2}(x)$ and $L^{(5)}_{d;17}$ has singularities at
$x=0,\infty,~1,~-1,~x_3=e^{\pm2\pi i/3},~x_5=e^{\pm 2\pi i/5},~e^{\pm
  4\pi i/5}$
where the non-integer exponents at $x_3$ are $5/2,7/2,7/2$ and at
$x_5$ are $23/2$.  It has been further  found that
\begin{equation}
L^{(5)}_{d;17}=L^{(5)}_{d;6}\cdot L^{(5)}_{d;11}
\end{equation}
with
\begin{equation}
L^{(5)}_{d;11}=L^{(3)}_{d;1}\oplus L^{(3)}_{d;3} \oplus
\left(W^{(5)}_{d;1}\cdot U^{(5)}_{d;1}\right) \oplus \left(L^{(5)}_{d;4}\cdot
V^{(5)}_{d;1} \cdot U^{(5)}_{d;1}\right)
\end{equation}
where $L^{(3)}_{d;m}$ annihilates $\chi^{(3)}_{d;m}$ and the remaining
 operators in this decomposition are all given in  \cite{mccoy11}.

\subsection{Singularities and cancellations}

By examining the integral representations for $\chi^{(n)}(w)$
(\ref{bernie}) and $\chi^{(n)}_{d}(t)$ (\ref{echidm}), (\ref{echidp})
it is clear that these integrals
have no singularities for $|\sinh 2E/k_BT|<1$. The
singularities at $|\sinh2E/k_BT|<1$ of the differential 
equations  for $\chi^{(n)}(w)$ will only appear in analytic
continuations of the integral in the complex plane of the variable
$w$. The corresponding differential equations for $\chi^{(n)}_d(t)$
are significantly simpler because they have singularities only 
at $t=0,\infty$ and $|t|=1$.

It remains to discuss the singularities in the differential equations
which do lie on   $|\sinh 2E/k_BT|=1$ and to give an explanation for
the observation that the singularities of the ODEs for
$\chi^{(n-2m)}(w)$ and $\chi^{(n-2m)}_d(t)$ are also singularities of
$\chi^{(n)}(w)$ and $\chi^{(n)}(t)$ respectively even though the
integrands are singular only at the points given by (\ref{location}) for
$\chi^{(n)}(w)$ and by (\ref{t0m}) and (\ref{t0p})  for
$\chi^{(n)}_d(t)$.

The resolution of this is easily seen for $\chi^{(n)}_d(t)$. By an
examination of the integrals (\ref{echidm}) and  (\ref{echidp}) we see that
there are paths of analytic continuation possible in the complex
 $t$ plane where the contour of integration must be deformed past the
pole at
\begin{equation}
1-tx_{2j}x_{2k+1}=0
\end{equation}
and the residue at that pole will reduce the denominators
in $\chi^{2n}_d(t)$ and $\chi^{(2n+1)}_d(t)$
from
\begin{equation}
1-t^nx_1\cdots x_{2n}
\end{equation}
and
\begin{equation}
1-t^{n+1/2}x_1\cdots x_{2n+1}
\end{equation}
to the denominators in $\chi^{(2n-2)}_d(t)$ and $\chi^{(2n-1)}_d(t)$
respectively with $n\rightarrow n-1$ and two less integration
variables. Therefore, the singularities of $\chi^{(n-2m)}_d(t)$ will
not appear on the principle sheet of the integral which is analytic at
$t=0$ but only on analytic continuations to non-physical branches.
The similar phenomenon occurs for $\chi^{(n)}(w)$.

It remains to reconcile this non appearance of the singularities of
$\chi^{(n-2m)}_d(t)$ in the physical sheet of $\chi^{(n)}_d(t)$ with
the direct sum decompositions (\ref{chidedsum}) and (\ref{chidodsum}).
This will be accomplished by showing that the term $\Omega^{(n)}_d(t)$
has singularities which exactly cancel the singularities on
$\chi^{(n)}_d(t)$. This requires the solution of a global connection
problem which has not yet been explicitly done even though from the
examination of the original integral the resulting exact cancellation
must hold.

\section{Conclusion}

Now that we have summarized the known features of the Ising
correlations, form factors and
susceptibility we can proceed to discuss what is not known. This is
the fascinating,  mysterious and thus romantic part of the subject.

\subsection{Conformal and quantum field theory}

One of the most important features of the Ising model is that the
scaling limit satisfies all the axioms for a massive Euclidean 
quantum field theory   
and that at $T=T_c$ the long range correlations are those of a 
conformal field theory with central charge
$c=1/2$. This is in fact the earliest conformal field theory known 
and from this beginning a vast new field of mathematics and physics
has been developed in the last 30 years. However, the Ising model is
much more than a conformal field theory because we have a vast number
of results for  $T\neq T_c$ which are the simplest examples of
properties of massive Euclidean quantum field theories. Part of the
romance is the exploration of how these Ising results can be used to  
extend massless conformal field theories into the massive region.

\subsection{Form factors, exponential forms and amplitudes}


The derivation \cite{mccoy5} of the exponential
and form factor expansion for the diagonal Ising correlation is much
more general than this special case. Indeed in \cite{mccoy5}
it is proven that every Toeplitz determinant (\ref{dn}) with a
generating function $\phi(\xi)$ such that $\ln \phi(\xi)$ is
 continuous and periodic on
$|\xi|=1$ has both an exponential and a form factor
expansion. Furthermore these Toeplitz determinants are also
expressible as Fredholm determinants \cite{bo} (at times in several different
ways \cite{wf}). Consequently the Ising computations  
have subsequently been extended to several very important problems
including the seminal work on the one dimensional impenetrable Bose
gas and on  random matrices by Jimbo, Miwa, Mori and
Sato\cite{jmms}.

To illustrate the differences between the form factor and the
exponential representation of the correlation functions, we consider
the computation by Tracy \cite{tracy} of the constant $A_{T_c}$ 
of (\ref{atc}). 
In the scaling limit the scaled correlations in
the general case where $E^v\neq E^h$ depend only on the single variable
$r$ (\ref{rdef}). Therefore we can restrict attention to the scaled form
of the diagonal correlation $\langle\sigma_{0,0}\sigma_{N,N}\rangle$
and consider the lambda extension of the scaling form 
of the exponential form (\ref{eformm})
which we write as
\begin{equation}
G_{-}(r;\lambda)=\exp\sum_{n=1}^{\infty}\lambda^{2n}g^{(2n)}(r),
\end{equation}
where  
\begin{equation}
g^{(2n)}(r)=\lim_{\rm scaling}F^{(2n)}_{N,N}(t),
\end{equation}
which depends on the single variable $r$ instead of the two independent
variables $N$ and $t$. Tracy finds that, as $r\rightarrow 0$, 
\begin{equation} 
g^{(2n)}(r)=-\alpha_n\ln r+\beta_n+o(1).
\end{equation}
Therefore, defining the lambda dependent sums
\begin{equation}
\alpha(\lambda)=\sum_{n=1}^{\infty}\lambda^{2n}\alpha_n
\hspace{.5in}\beta(\lambda)=\sum_{n=1}^{\infty}\lambda^{2n}\beta_n,
\label{ablam}
\end{equation}
we find
\begin{equation}
G_{-}(r;\lambda)\sim\exp\{-\alpha(\lambda)\ln r+\beta(\lambda)\}
=\frac{e^{\beta(\lambda)}}{r^{\alpha(\lambda)}}.
\label{grl}
\end{equation}
In the Ising case where $\lambda=1$ the functions specialize to
\begin{equation}
\alpha(1)=1/4\hspace{.5in}\beta(1)=\ln A
\end{equation}
where $A$ is the constant in (\ref{ctc}).

If, however, instead of the scaled exponential form we define the
scaled limit of the form factors $f^{(2n)}_{N,N}(t)$ as
\begin{equation}
{\tilde f}^{(2n)}(r)=\lim_{\rm{scaling}}f^{(2n)}_{N,N}(t),
\end{equation}
then as $r\rightarrow 0$
\begin{equation}
{\tilde f}^{(2n)}(r)=\sum_{k=0}^{n}a^{(2n)}_k \ln^k r +o(1).
\label{tone}
\end{equation}
Thus, in order for (\ref{grl}) to agree with the $r\rightarrow 0$
behavior of the form factor expansion, we need
\begin{eqnarray}
\frac{e^{\beta(\lambda)}}{r^{\alpha(\lambda)}}
&&=
\left[\sum_{k=0}^{\infty}\frac{1}{k!}
\left(\ln r\sum_{n=1}^{\infty}\lambda^{2n}\alpha_n\right)^k\right]
\left[\sum_{k=0}^{\infty}\frac{1}{k!}
\left(\sum_{n=1}^{\infty}\lambda^{2n}\beta_n\right)^k\right]\nonumber\\
&&=1+\sum_{n=1}^{\infty}\lambda^{2n}\sum_{k=0}^na_k^{(2n)}\ln^k r
\label{id}
\end{eqnarray}
to hold term by term for each power $\lambda^{2n}$. The requires an
infinite number of identities between the $a^{(2n)}_k$.


As an additional remark we note that if
we rewrite the integral (\ref{fdm}) for $f^{(2n)}_{N,N}$ as a contour
integral, rescale the variables $x_k$ by $x_k=t^{-1/2}y_k$
and then send $y_{2k}\rightarrow 1/y_{2k}$ we see that as
$t\rightarrow 1$ the integral has logarithmic divergences as in
(\ref{tone}). The amplitudes $a_n$ are closely related to the
special case with $\rho=1$ of the integral found by Dotsenko 
and Fateev \cite{df}  in their study of four point correlations 
in conformal field theories with central charge $c\leq 1$
\begin{eqnarray}
I_{n,m}(\alpha,\beta;\rho)&&=\frac{1}{n!m!}\prod_{i=1}^n\int_0^1 
dt_i t_i^{\alpha'}(1-t_i)^{\beta'}
\prod_{i=1}^m\int_0^1d\tau_i^{\alpha}(1-\tau_i)^{\beta}\nonumber\\
&&\times|\prod_{i<j}(t_i-t_j)|^{2\rho'}|\prod_{i<j}(\tau_i-\tau_j)|^{2\rho}
\prod_{i,j}^{n,m}\frac{P}{(t_i-\tau_j)^2}
\end{eqnarray}  
where $P$ indicates the principal value and
\begin{equation}
\alpha'=-\rho'\alpha,~~~\beta'=-\rho'\beta,~~~\rho'=\rho^{-1}
\end{equation}

\subsection{Exponentiation}

Form factor expansions exist for many massive models of quantum
field theory including sine-Gordon and the non-linear sigma model 
\cite{smirnov} and similar form factor expansions exist 
\cite{maillet1},\cite{maillet2}  for the
XXZ model on a chain of finite length 
\begin{equation}
  H_{XXZ}=-\sum_{j=1}^L\{\sigma_j^x\sigma_{j+1}^x+
\sigma_j^y\sigma_{j+1}^y+\Delta\sigma_j^z\sigma_{j+1}^z+H\sigma_j^z\}
\end{equation}
Moreover the Feynman expansion of amplitudes in quantum field theory is
also what we have called here a form factor expansion. In all of the
models there are limiting cases where series of multiple dimensional
integrals expand to series in powers of logarithms which need to be
summed.  
However, unlike the Ising correlation functions these form
factor expansions do not come from either Toeplitz or Fredholm
determinants and thus the exponentiation methods of Ising correlations
are not applicable.

Over the years an immense effort has been made to sum the form factor
series of logarithms. In quantum field theory this starts with the
classic 1939 paper of Bloch and Nordsiek \cite{bn} on resummation of
infrared divergences in quantum electrodynamics. 
A second example is the Regge theory of the 60's and 70's 
 an 2nth order  Feynman diagram expansion of a four point scattering 
amplitude is shown to diverge as the energy $s\rightarrow \infty$
with a fixed momentum transfer $t$ as
\begin {equation}
g^{2n}\alpha(t)^n\frac{\ln^n s}{n!}
\end{equation}
This is a ``leading'' log approximation and is analagous to the Ising
case if only the first term in the series (\ref{ablam}) 
for $\alpha_n$ is retained. 
More recently there has been a great deal of work on quantum
chromodynamics \cite{qcd} where many non leading terms summed by the
use of an ingenious decomposition of the multidimensional integrals.

In the theory of integrable systems a great
deal of effort has been devoted to compute the long range 
asymptotic behavior of the correlations of the XXZ model 
in the massless region $-1<\Delta <1$ 
from multiple integral representations. One 
method is presented in \cite{maillet3} which shows how to modify the
Fredholm determinant form which holds for $\Delta=0$ by suitably
picking out the important pieces of the multiple integrals. This has
led to the computation of both the exponents and the amplitude of the
long range behavior of the correlations when the $H\neq 0$. The study
of the  correlations from the form factors is begun in
\cite{maillet1}-\cite{maillet2} with more results announced to be
forth coming. A full exploration of the relation of these  subjects
is beyond the scope of this article.

\subsection{Short distance versus scaling terms}

In the $n$-particle expansions of the full (\ref{chim}), (\ref{chip}) and
the diagonal (\ref{chidm}), (\ref{chidp}) susceptibility
the $\chi^{(n)}(t)$ and the $\chi^{(n)}_d(t)$ will (for $n \geq 3$)
have terms which contain powers of $\ln t$. From this it might be
inferred that the susceptibility will contain terms of the form
$(1-t)^{1/4+p}\ln^q(1-t)$. However, from the extensive calculations on
long low and high temperature series expansions made in \cite{ongp}
and \cite{cgnp} such terms do not appear to exist. Instead the
susceptibility is conjectured to have the form for $t\rightarrow 1$ of
\begin{equation}
k_BT \chi(t)_{\pm}=(1-t)^{-7/4}\sum_{j=0}^{\infty}C^{(j)}_{\pm}(1-t)^j
+\sum_{q=0}^{\infty}\sum_{p=0}^{[\sqrt q]}b_{\pm}^{(p,q)}(1-t)^q\ln^p(1-t) 
\end{equation}
The first term is called the ``scaling function''.
The second term is called the ``background'' or ``short distance'' 
term and is numerically obtained by summing correlation functions 
instead of form factors. In \cite{cgnp} it is stated that the ``scaling  
function'' is determined by conformal field theory while for the
``short distance'' term here is ``no explicit prediction''. In
\cite{ongp} the belief is stated that the separation into ``scaling''
and ``short distance'' parts is ``tantamount to the scaling argument
that in the critical region there is a single length scale proportional
to $(1-t)^{-\nu}$ with $\nu=1$''. It would be highly desirable if this
distinction between ``scaling'' and ``short distance'' terms could be
made precise and if both terms could be obtained by use of  the form factor 
expansion alone.

\subsection{Natural boundaries and $\lambda$ extensions}

Perhaps the most perplexing question concerning the relation of the
Ising model on a lattice with the scaling field theory limit is the
existence of the natural boundary in the susceptibility implied by the
singularities (\ref{singodd}), (\ref{singeven}) 
found by Nickel \cite{ni1}, \cite{ni2}.
The magnetic susceptibility is the second derivative of the free
energy with respect to an external magnetic field $H$ interacting with the
spins as $-H\sum_{j,k}\sigma_{j,k}$. In the scaling limit 
the Ising model in a magnetic field is also a field theory  and the
analyticity properties of this field theory have been extensively
studied by Fonseca and Zamolodchikov \cite{fz} with the conclusion 
that there is no natural boundary. How can this be reconciled with the
computations of \cite{ni1} and \cite{ni2}? 

The existence of the natural boundary suggested by Nickel in \cite{ni1} and
\cite{ni2} rests on the accumulation of the 
singularities (\ref{singodd}) and (\ref{singeven}) and the assumption
that there is no cancellation. However, for this argument to hold we
need to be able to show that the limit of $t$ approaching the
location of the supposed natural boundary (\ref{location})
will commute with the infinite sum over the $n$ particle
contributions $\chi^{(n)}(T)$ in (\ref{chip}) and (\ref{chim}). 
Since the natural boundary does not exist if only a finite number of the 
$\chi^{(n)}(T)$ are included this interchange need to be
investigated. It is also possible that the existence of a natural
boundary could depend on the value of $\lambda$ in the lambda
extensions of (\ref{chip}) and (\ref{chim})
\begin{eqnarray}
k_BT\chi_+(T;\lambda)
&=&(1-t)^{1/4}t^{-1/4}\sum_{j=0}^{\infty}\lambda^{2j}{\chi}^{(2j+1)}(T)
~~{\rm for}~ T>T_c\label{fullp}\\ 
k_BT\chi_-(T;\lambda)&=&(1-t)^{1/4}\sum_{j=1}^{\infty}\lambda^{2j}{\chi}^{(2j)}(T)
~~{\rm for}~T<T_c.\label{fullm}
\end{eqnarray}
These possibilities remain to be investigated.

\subsection{Row correlations}

All of the results obtained for the diagonal
correlation, which depend on the single variable $t$, can be extended to
the row correlation, which depends on the two variables $\alpha_1$ and
$\alpha_2$ in a symmetric fashion (\ref{an}). In particular it has
been pointed out to me by Jean-Marie Maillard and Nicholas Witte in
private conversations that the Painlev{\'e} VI results of Jimbo and
Miwa \cite{jm} can be extended to a two variable Garnier system
\cite{gar}. However, this system must possess some most interesting properties
because one of the most important properties of the Ising model is the
fact that, when these two variables are rewritten as
\begin{equation}
k=\sinh 2E^v/k_BT\sinh 2E^h/k_BT~~{\rm and}~~r=\frac{\sinh
  2E^v/k_BT}{\sinh 2E^h/k_BT},
\end {equation}
the dependence on $k$ (the modulus of the elliptic functions) 
and the anisotropy ratio $r$ which is related to the spectral variable
of the star triangle equation \cite{baxb} is dramatically different.
These results for Garnier systems have also yet to be obtained.

\section{Romance versus Understanding}

In a lecture given in Melbourne in January 2006 \cite{mccoy10},
I gave the following definition of ``understanding''

\vspace{.1in}

No one can be said to understand a paper unless he is able to
generalize the paper.

\vspace{.1in}

This definition is open to criticism on at least two grounds. Firstly the use
of the word ``he'' has a sexist implication which is neither
appropriate nor intended. Secondly, there are surely subjects which
are fully understood where further generalization is pointless. An
illustration of this are the laws of thermodynamics which have been
fully understood by physicists for many decades (even if they are not
accepted by the overwhelming majority of voters and politicians).

However, precisely  because thermodynamics is fully understood, 
it has lost the mystery it had at the time of Gibbs, Boltzmann and
Ehrenfest. This illustrates the great truth that 
understanding is the enemy of romance because once the mysteries are
understood the romance dies.

Fortunately for romance, there are many mysteries of the Ising model 
which are far from being understood.
The romantic in me says that, even when these mysteries have been
understood, the understanding of the mysteries will generate new
mysteries and the romance of the Ising model will  be
everlasting.

{\bf Acknowledgments}

In my long running romance with the Ising model I have profited
greatly from the help many people. In particular I
want to thank M. Assis, H. Au-Yang, R.J. Baxter, V.V. Bazhanov, P.J. Forrester,
 M. Jimbo, J-M. Maillard, J-M. Maillet, T. Miwa, W. Orrick, 
J.H.H.Perk, C.A.Tracy, N. Witte,  and 
T.T. Wu for their wisdom and inspiration.

\end{document}